# Multimode Vibrational Strong Coupling of Methyl Salicylate to a Fabry-Pérot Microcavity

Wassie Mersha Takele,[1,2] Frank Wackenhut,[2,*] Lukasz Piatkowski,[1,3,*] Alfred J. Meixner,[2] and Jacek Waluk[1,4]

[1]*Institute of Physical Chemistry, Polish Academy of Sciences, Kasprzaka 44/52, 01-224 Warsaw, Poland;*
[2]*Institute of Physical and Theoretical Chemistry and LISA+, University of Tübingen, Auf der Morgenstelle 18, D-72076 Tübingen, Germany;*
[3]*Faculty of Technical Physics, Poznan University of Technology, Piotrowo 3, 60-965 Poznan, Poland;*
[4]*Faculty of Mathematics and Science, Cardinal Stefan Wyszyński University, Dewajtis 5, 01-815 Warsaw, Poland;*
*Authors contributed equally: frank.wackenhut@uni-tuebingen.de; lukasz.j.piatkowski@put.poznan.pl

**Abstract**

The strong coupling of an IR-active molecular transition with an optical mode of the cavity results in vibrational polaritons, which opens a new way to control chemical reactivity via confined electromagnetic fields of the cavity. In this study, we design a voltage-tunable open microcavity and we show the formation of multiple vibrational polaritons in methyl salicylate. A Rabi splitting and polariton anticrossing behaviour is observed when the cavity mode hybridizes with the C=O stretching vibration of methyl salicylate. As this vibration contributes to the reaction coordinate of the photoinduced proton transfer process in methyl salicylate, we suggest the coupling might be used to modulate the photophysical properties of the molecule. Furthermore, the proposed theoretical model based on coupled harmonic oscillator reveals that the absorption of uncoupled molecules must also be considered to model the experimental spectra properly and that simultaneous coupling of multiple molecular vibrations to the same cavity mode has a significant influence on the Rabi splitting.

**TOC Graphic**

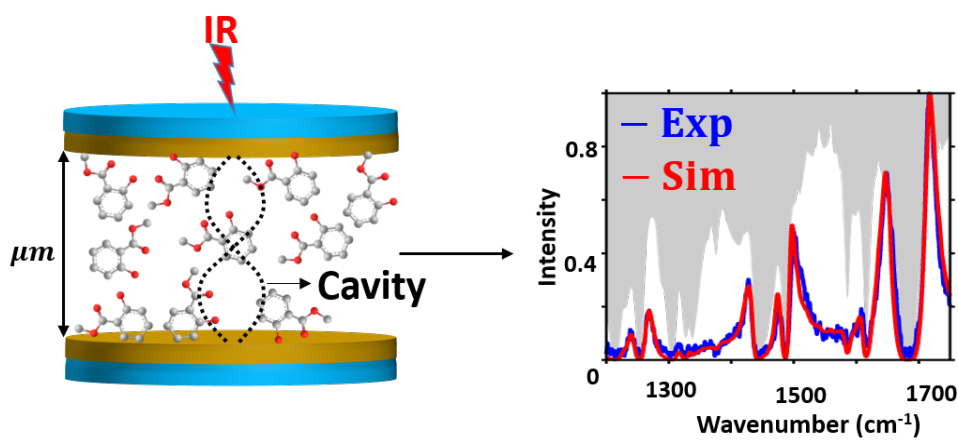

**Keywords:** Strong coupling, polaritons, Rabi splitting, molecular vibrations, proton transfer.

When molecules are placed within an optical microcavity, their quantized transitions can exchange energy with the cavity mode once the resonant condition is met.[1,2] If the rate of energy exchange is faster than the decay rates of both constituents, the system enters the so-called strong coupling (SC) regime.[3] SC leads to the formation of hybrid light-matter states called polaritons.[4,5] Since this opens a way to tune energy levels,[6,7] SC has emerged as an intriguing platform to actively control molecular and material properties.[8–15] For instance, strong coupling of electronic excited states with the cavity mode has been used to modulate energy transfer,[16] conductivity,[17] Stokes shift,[18] quantum yield,[19] and work function.[20] Recently, these ideas have been extended to a novel field of vibrational strong coupling (VSC), where chemical reactivity can be altered by the zero-point energy fluctuations of the optical mode of a cavity.[21–23]

VSC is attained when IR-active molecular vibrations hybridize with the vacuum electric field of an IR- microcavity.[24] As a consequence, a vibrational transition splits into lower (VP$^-$) and higher (VP$^+$) vibrational polaritons,[25] separated by the Rabi splitting energy ($\hbar\Omega_R$), which is illustrated in Figure 1a. The splitting depends on the molecular transition dipole moment ($\mu$), the cavity electric field (E$_{cav}$) and the concentration (C), $\hbar\Omega_R \propto \mu \cdot \mathrm{E_{cav}} \cdot \sqrt{\mathrm{n_{ph}}+1} \cdot \sqrt{C}$, where n$_{ph}$ is the number of photons in the cavity.[6,26] Vibrational polaritons have been observed from polymer films,[27–29] pure organic liquids,[26] organometallic complexes dissolved in water,[30] a liquid crystal molecule,[25] and different molecules simultaneously dissolved in solution.[31] So far, VSC has been employed to supress[32] or enhance[33] the rate of chemical reactions and even alter reaction selectivity.[34] Recently, Vergauwe et al. showed the modification of the rate of an enzymatic hydrolysis reaction under strong coupling of the water vibrations.[35] An intriguing idea is to alter/control intramolecular reactions to tailor the photophysics and/or excited state reactivity of a molecule. Fascinating examples are molecules exhibiting excited state intramolecular proton transfer, for instance methyl salicylate (MS), see Figure 1b for the molecular structure, which we selected as a model system for our studies. It has been demonstrated for several proton transferring systems that different vibrational modes may contribute to the tautomerization coordinate.[36-38] One can therefore expect that shifting the energy levels of these modes by coupling to the cavity may influence the reaction. Thus, as a prelude to controlling photoinduced tautomerization through VSC, we investigate here the properties of MS in the strong coupling regime.

In most of the previous studies,[26-31] VSC is typically realized using a fixed thickness microcavity, for which the resonant frequency is tuned by tilting the sample with respect to the light propagation direction. However, in this work, we made use of a voltage-tunable open Fabry-Pérot microcavity that can be brought into resonance with any vibrational transition without moving and/or rotating the sample, thereby it could be an important tool to the growing field of VSC. Coupling of two vibrational frequencies from a single molecular species with a cavity mode was first reported by George et al.,[26] while Menghrajani et al.[39] recently demonstrated the hybridization of three vibrational resonances of a polymer film with the same cavity mode. However, the effect of simultaneous coupling on the Rabi splitting has not been explored. Here, we show that multiple vibrations from MS solution, even those seemingly OFF-resonant, can be simultaneously hybridized to the same cavity mode. To examine the influence of this complex multimode coupling behaviour on the Rabi splitting we develop a theoretical model based on a damped harmonic oscillator.

The design of the piezo-based tunable Fabry-Pérot open microcavity is presented in Figure 1c.[40-45] In a nutshell, the cavity consists of two Au-coated $CaF_2$ windows (see Supporting Information). The two mirrors are mounted into mirror holders and the optical path length ($L_{op}$) of the cavity is tuned by applying a voltage to the piezoelectric stacks integrated in one of the mirror mounts. The exemplary transmission spectra of an empty microcavity as a function of applied voltage are shown in Figure 1d. For each voltage increment (1V), 5 cm$^{-1}$ shift of the resonant frequency is observed. A typical Q-factor of 70-100 was achieved and the optical path length was varied ($L_{op}$ ~12 µm - 30 µm) depending on the concentration of the sample, yielding the FSR in the range of 200-400 cm$^{-1}$. The microcavity employed in this work has the advantage that the parallelism of the cavity mirrors can still be controlled after assembly, along with a nearly unlimited tuning range and can be used for coupling of the cavity mode with essentially any molecular vibration in the mid infrared region.

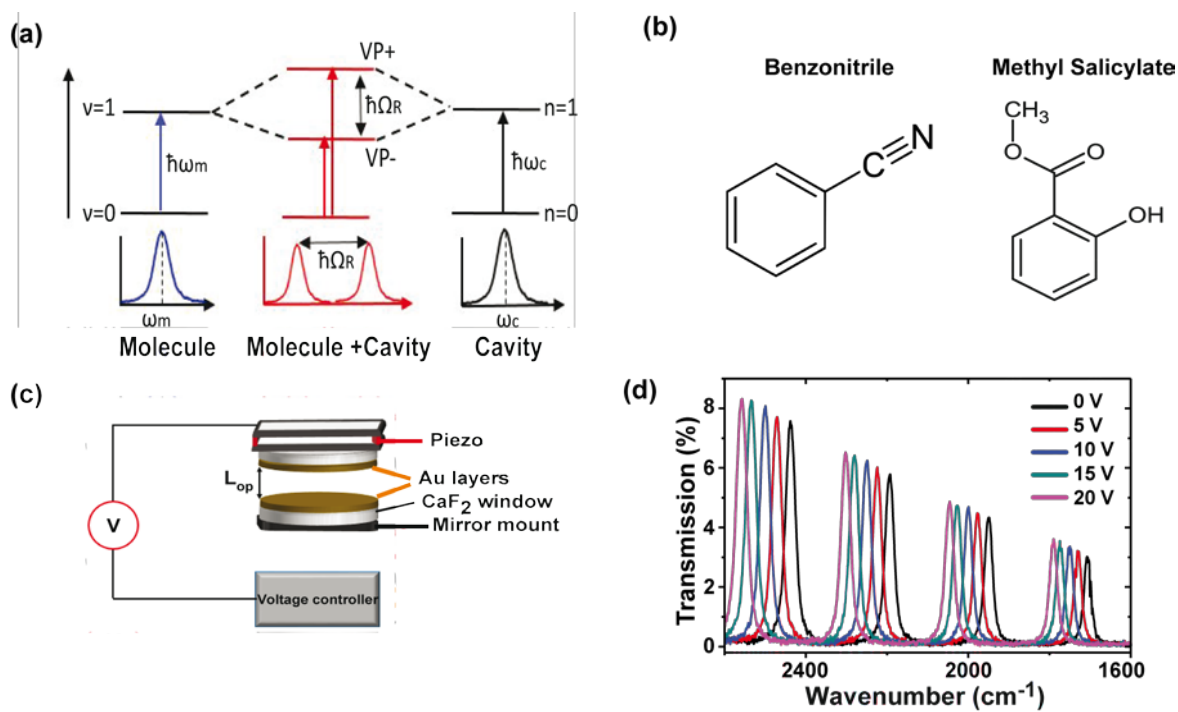

**Figure 1** (a) Schematic illustration of the strong coupling effect between a vibrational transition (v=0 → v=1) and an optical microcavity mode (n=0 → n=1). When the vibrational frequency of a molecular transition ($\omega_m$) matches the resonant frequency of a cavity mode ($\omega_c$), the two states hybridize and split into two new states: lower (VP⁻) and upper (VP⁺) vibrational polaritons, separated by the Rabi splitting ($\hbar\Omega_R$). (b) Molecular structure of benzonitrile and methyl salicylate. (c) Schematic illustration of the Fabry-Pérot type open microcavity structure used to couple liquid phase samples. (d) Transmission spectrum of an empty microcavity, as a function of voltage applied to a piezoelectric element of the mirror holder.

The FTIR spectrum of MS diluted in methylcyclohexane is shown in black in Figure 2a. It exhibits an intense C=O stretching frequency at 1685 cm$^{-1}$. The vibrational frequency of free aromatic C=O ester resides usually between 1750 and 1735 cm$^{-1}$.[46] In MS hydrogen bonding leads to a redshift of the C=O stretching frequency, which is consistent with values reported in the literature.[47, 48] The bands observed at 1585 cm$^{-1}$ and 1616 cm$^{-1}$ in the FTIR spectrum of MS arise from C=C vibrational stretching modes of the phenyl moiety.[49]

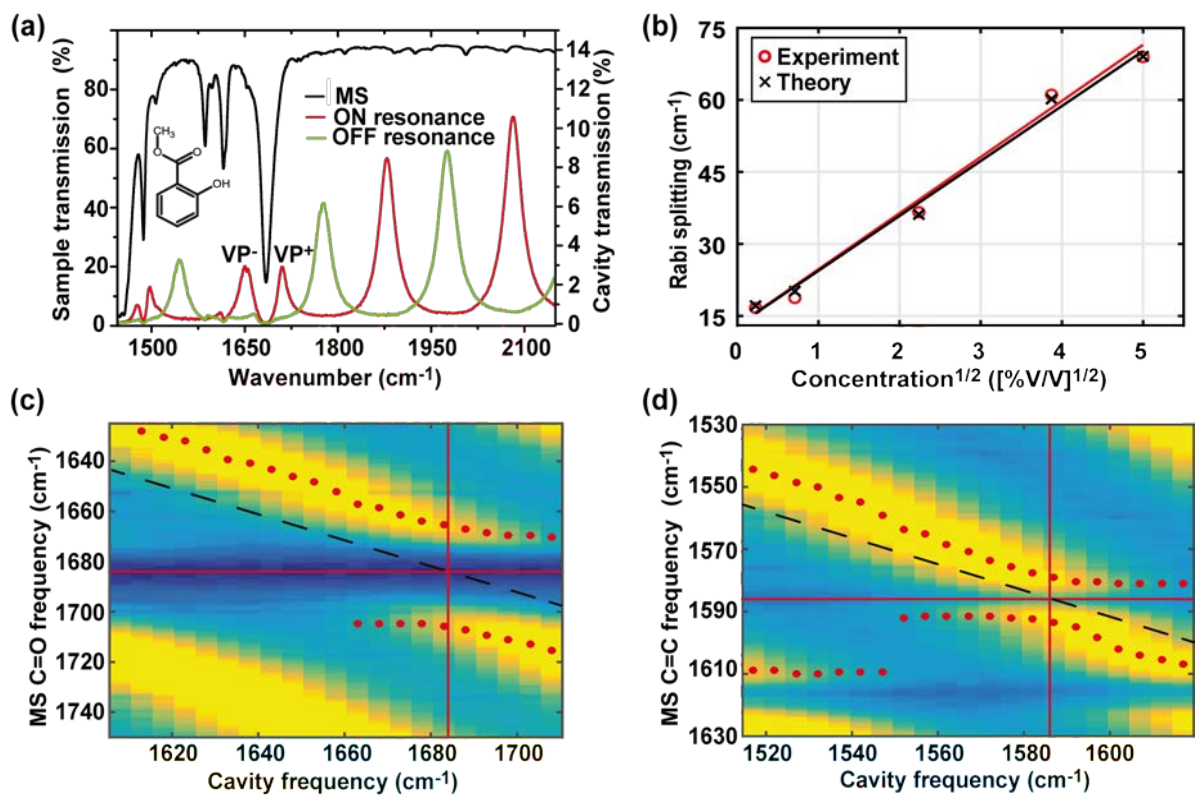

**Figure 2** (a) FTIR transmission spectrum of 15% v/v MS dissolved in methylcyclohexane (black curve). The transmission spectrum of a microcavity filled with MS solution, ON (red) and OFF (green) resonance, respectively. (b) Concentration dependence of the Rabi splitting of the C=O stretching mode of MS. Experimental and calculated data are shown in red and black, respectively, and are obtained from the spectral separation of the VP⁻ and VP⁺. The solid lines are linear fits to the data. The calculated results are taking into account simultaneous coupling of three vibrations to the same cavity mode. The dotted line indicates the transition between the weak and strong coupling regimes. (c-d) Dispersion plots for the C=O (c) and C=C (d) vibrations of MS (5% v/v) coupled to a microcavity. Dots indicate maxima of the intensity of polaritonic peaks. Solid red lines indicate resonant frequencies of the cavity and C=O and C=C stretching modes, while black dashed lines indicate the cavity mode tuning, highlighting the anti-crossing character of the dispersion plot.

In order to create the vibrational polaritonic states in MS, the cavity was tuned towards the C=O vibrational band. ON resonance two new transitions, at 1650 cm⁻¹ and 1710 cm⁻¹, are observed in the transmission spectrum of the coupled system (Figure 2a, red curve). These are the signatures of the vibrational hybrid light-matter states (VP⁺ and VP⁻) in MS. In the OFF resonance case (Figure 2a, green curve) the VSC effect is not observed, yet a minute dispersive line shape is detected at the vibrational mode frequency. The separation between VP⁺ and VP⁻ (60 cm⁻¹) - the Rabi splitting - is larger than the linewidth of the cavity mode (30 cm⁻¹) and of the molecular resonance (16 cm⁻¹). Since the coupling exceeds the linewidths of both resonator and molecule, the hybridization of the cavity mode and the C=O transition in MS satisfies the criterion of the strong coupling.[14]

Changing the concentration of MS in the mode volume of the cavity leads to an increase of the observed Rabi splitting, which is shown experimentally in red in Figure 2b. The dashed lines are linear fits to the data. At the lowest concentrations (0.05% v/v and 0.5% v/v), the values of the Rabi splitting are smaller than the bandwidth of the cavity mode. However, when the concentration increases to 5% v/v, the value of the splitting energy exceeds the bandwidth of the cavity and the coupled system passes from weak to strong coupling regime. The black crosses display the peak separation calculated with a harmonic oscillator approach, which is discussed below and in the Supporting Information. Figures 2c-d display the polariton anticrossing behaviour as a function of the cavity tuning. The bare C=O vibration and the bare cavity mode cross each other (marked by two red lines in Figure 2c), while the maxima of the VP$^+$ and VP$^-$ show avoid crossing, which is yet another signature of the strong coupling in MS.

The theoretical description of VSC is based on coupling of up to four damped harmonic oscillators and is described in the Supporting Information and by Junginger et al.[50] Benzonitrile (BN) was used as a simple model compound, since it has a well-isolated C≡N vibration. Figure 3a-b show the experimental (blue line) and simulated (red line) transmission spectrum of a cavity containing neat benzonitrile. The dashed vertical lines indicate the three cavity resonances. The grey area illustrates the free space transmission spectrum of a BN solution.

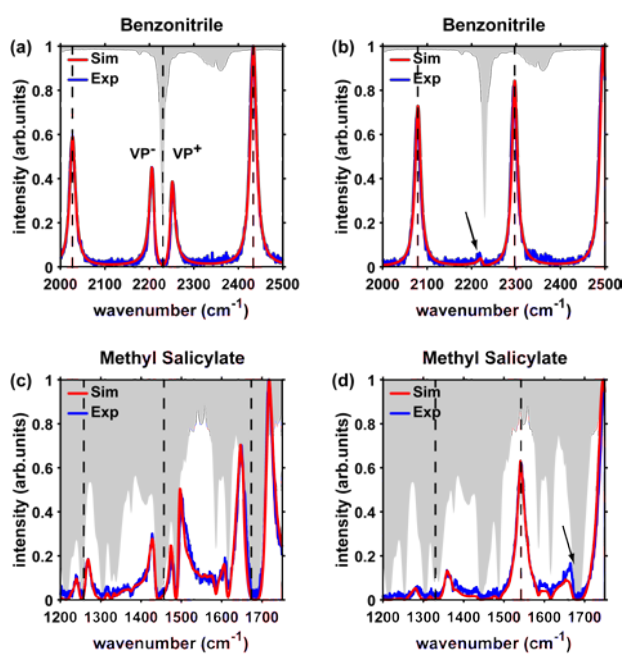

Figure 3 (a)/(b) Experimental (blue line) and simulated (red line) cavity transmission spectra. The cavity is ON resonance in (a) and OFF resonance in (b) with the $C \equiv N$ vibration of benzonitrile. A splitting into the lower (VP$^-$) and upper (VP$^+$) vibrational polaritons is clearly observable. The free space transmission spectrum is indicated by the grey area. (c)/(d) Experimental (blue) and simulated (red line) transmission spectra of methyl salicylate in a resonant (c) and an OFF resonant cavity (d). The transmission spectrum exhibits a complex mode coupling. The transmission spectrum of MS solution (25% v/v) is indicated by the grey area. The dashed lines indicate the cavity modes.

In Figure 3a splitting into the lower (VP$^-$) and upper (VP$^+$) vibrational polariton can be observed in the transmission spectrum when the $C \equiv N$ stretching mode at 2229 cm$^{-1}$ is hybridized with the cavity mode. The separation between the VP$^+$ and VP$^-$ peaks is 50 cm$^{-1}$, which is comparable with values reported by George et al.[26] The experimental (blue line) and simulated (red line) transmission spectra of the same cavity are shown in Figure 3b, but in this case the cavity resonance is detuned from the $C \equiv N$ vibration. The transmission spectrum is similar to an empty cavity and only a weak signature of coupling is observed (marked with an arrow). We found a very good agreement between the experimental and simulated data, which shows that the coupled oscillator approach is suitable to model the strongly coupled molecule-cavity hybrid system. Figures 3c and 3d show an equivalent experiment with MS in an ON and OFF resonant cavity, respectively. When one of the cavity resonances is tuned towards the C=O vibrational band at 1685 cm$^{-1}$, the Rabi splitting is observed in the transmission spectrum of the coupled system (Figure 3c). Two other cavity resonances are located at 1275 cm$^{-1}$ and 1465 cm$^{-1}$ and the mode splitting is more complex due to coupling of multiple vibrations to the same cavity mode. Figure 3d shows the cavity transmission spectrum for another cavity length, where one cavity mode at 1323 cm$^{-1}$ couples to vibrations in this spectral range, while two cavity modes at 1533 cm$^{-1}$ and 1730 cm$^{-1}$ are off resonant to the strongest vibrations. Nevertheless, coupling to the intense C=O stretching vibration can still be observed at 1685 cm$^{-1}$ (indicated by the arrow in Figure 3d), even though the cavity is not resonant to this vibration. Such an OFF resonant coupling is observed for various vibrations, e.g. at 1458 cm$^{-1}$, 1616 cm$^{-1}$, and is characterized by weak dispersive line shapes in the cavity transmission spectrum.

The transmission spectrum of a cavity containing MS is obviously more complex than for BN, since MS has a richer vibrational spectrum and multiple vibrations are spectrally close enough to couple to the same cavity mode. In the following, we investigate this coupling behaviour in detail. Figure 4 shows the experimental (blue lines) transmission spectra of a 5% v/v (top row) and 25% v/v (lower row) solution in a resonant cavity together with the corresponding simulations (red lines). The transmission spectrum of a bare MS solution with the same concentration is given by the grey area.

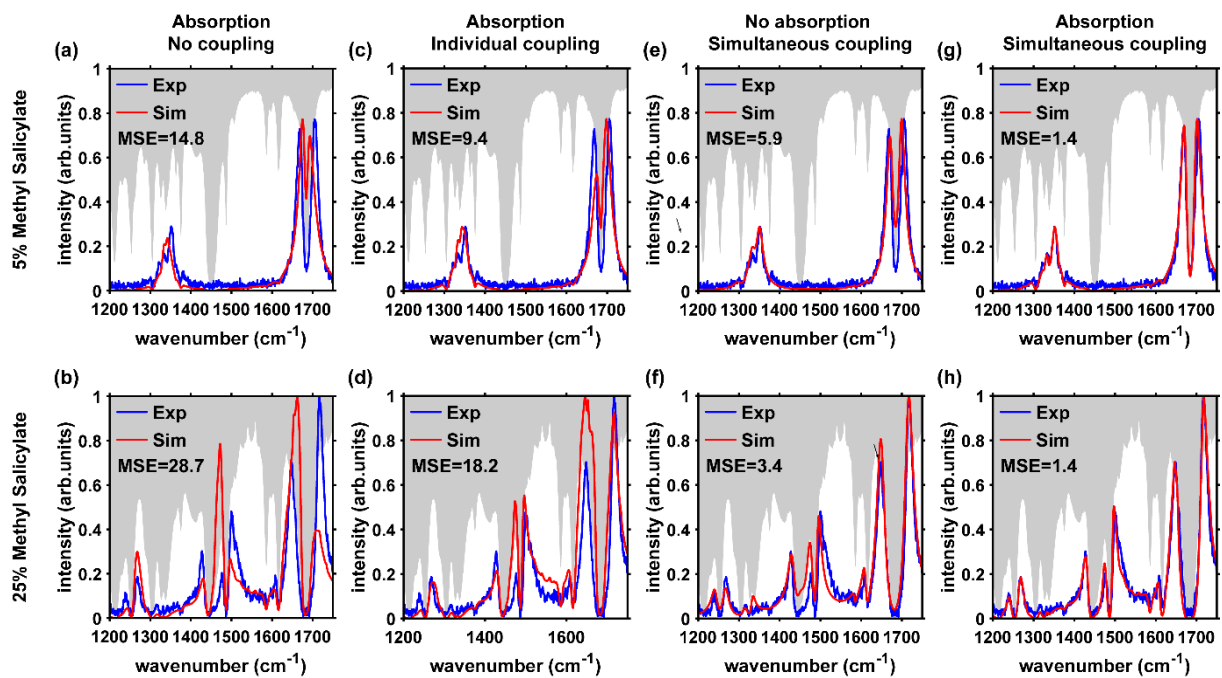

Figure 4: (a)/(b) Experimental (blue line) cavity transmission for a 5% v/v (a) and a 25% v/v (b) solution of MS inside an ON resonant cavity. The respective transmission spectrum of MS in free space is always indicated by the grey area. MSE indicates the mean square error and is a measure of the fit quality. The calculated spectrum shown in red only includes the absorption of MS and no VSC is considered. In (c)/(d) the same experimental (blue line) transmission spectra are shown, but are compared to simulations (red line) taking into account the absorption of the sample and strong coupling of the vibrations that are resonant with the respective cavity mode. (e)/(f) illustrate the case when in addition to resonant coupling of the cavity with a molecular mode simultaneously also the closest two off-resonant molecular vibrations are allowed to couple to the cavity mode. Finally, the simulations in (g)/(h) include absorption and simultaneous coupling of vibrations, which are spectrally close to the cavity mode.

Figure 4 compares four simulated coupling scenarios with the experimental transmission spectra of an ON resonant cavity. In the first case (Figure 4a/b) coupling of MS to the cavity mode is switched off in the simulations and only absorption of uncoupled MS molecules is considered. This is achieved by calculating the cavity transmission spectrum and multiplying it with the free space MS absorption spectrum and already leads to a splitting of the originally Lorentzian shaped cavity transmission spectrum, even though there is no VSC. However, there are various discrepancies between the calculated and the experimental spectrum, especially the dip at 1685 cm$^{-1}$ and the intensity ratios of all peaks are not reproduced well, leading to large mean square errors (MSE) of 14.8 and 28.7, respectively. Evidently, considering only absorption of uncoupled molecules is not sufficient to model the experimental transmission spectrum. In the second case (Figure 4c/d) we take both, absorption and VSC into account, but different MS vibrations couple individually to the cavity modes. Similar to Figure 4a/b we find a splitting of the cavity transmission spectrum into several modes, but the intensity ratios of the modes at 1465 cm$^{-1}$ and 1685 cm$^{-1}$ are not fitting the experimental data. Nevertheless, the MSE reduces to 9.3 and 18.2. The last two cases illustrate the effect case when in addition to resonant coupling of the cavity with a molecular mode simultaneously also the closest two off-resonant molecular vibrations are allowed to couple to the respective cavity mode without (Figure 4e/f) and with (Figure 4g/h) inclusion of absorption of uncoupled molecules. The agreement between the experimental data and the simulation increases when simultaneous coupling of the cavity mode with vibrations, which are spectrally close (see Figure S1 and S2 for assignment of the vibrations), is considered and the MSE decreases to 5.9 and 3.4, respectively. However, we find the best match between the simulation and the experimental data when simultaneous coupling and absorption is taken into account (Figure 4g/h) with a MSE of 1.4 for both concentrations. It is evident that absorption of uncoupled molecules needs to be considered and that we observe simultaneous hybridization of, at least, three MS vibrations with the same cavity mode. Finally, we can compare the experimental Rabi splitting with the corresponding calculations, which is shown in Figure 2b. The Rabi splitting of the experimental spectrum, obtained from the spectral separation between VP$^-$ and VP$^+$ peaks, is shown in red in Figure 2b. The black crosses in Figure 2b show the peak separation in the calculated spectra and there is an excellent agreement with the experimental values.

In summary, the open Fabry-Pérot microcavity used in this work is an appropriate tool for the emerging fields where vibrational strong coupling plays a central role as it can be fine-tuned to any molecular vibration. We found that in addition to resonant coupling of a molecular vibration to a cavity mode also coupling of the closest off-resonance molecular vibrations has to be taken into account for a good reproduction of the experimental spectra. This is an important observation, as it suggests that coupling, and therefore energy transfer, occurs even for molecular modes that are off-resonant but close to a cavity resonance. We showed that the complex coupling pattern between multiple molecular transitions with a single cavity mode is well captured by the coupled harmonic oscillators model. The best correspondence between the theory and experiment is found when the effects of absorption of uncoupled molecules and simultaneous coupling of close off-resonant modes are taken into account.

**AUTHOR INFORMATION**

Corresponding Authors
*E-mail: frank.wackenhut@uni-tuebingen.de (F.W.)
*E-mail: lukasz.j.piatkowski@put.poznan.pl (L.P.)

ORCID

Frank Wackenhut: 0000-0001-6554-6600

Lukasz Piatkowski: 0000-0002-1226-2257

Jacek Waluk: 0000-0001-5745-583X

Notes

The authors declare no competing financial interest.

**ACKNOWLEDGEMENTS**

W.M.T. acknowledges support from the European Union's Horizon 2020 research and innovation programme under the Marie Skłodowska-Curie grant agreement No. 711859 (CO-FUND NaMeS project) and the financial resources from Ministry of Science of Poland for science in the years 2017–2021 awarded for the implementation of an international co-financed project. The project has received funding from the National Science Centre, Poland, grants 2015/19/P/ST4/03635 and 2017/26/M/ST4/00872, POLONEZ 1 and 2017/26/M/ST4/00072, and from the European Union's Horizon 2020 research and innovation programme under the Marie Skłodowska-Curie grant agreement No. 665778. F.W. and A.J.M acknowledge support from the DFG grand ME 1600/13-3.

**Supporting Information Available**: Solution preparation, microcavity preparation, optical measurements, details of the theoretical model, methyl salicylate vibrational frequency peak assignments, and parameters used to simulate the spectra of benzonitrile and methyl salicylate.

# Supporting Information

# Multimode Vibrational Strong Coupling of Methyl Salicylate

# to a Fabry-Pérot Microcavity

Wassie Mersha Takele,[1,2] Frank Wackenhut,[2,*] Lukasz Piatkowski,[1,3,*] Alfred J. Meixner,[2] and Jacek Waluk[1,4]

[1]*Institute of Physical Chemistry, Polish Academy of Sciences, Kasprzaka 44/52, 01-224 Warsaw, Poland;*
[2]*Institute of Physical and Theoretical Chemistry and LISA+, University of Tübingen, Auf der Morgenstelle 18, D-72076 Tübingen, Germany;*
[3]*Faculty of Technical Physics, Poznań University of Technology, Piotrowo 3, 60-965 Poznań, Poland;*
[4]*Faculty of Mathematics and Science, Cardinal Stefan Wyszyński University, Dewajtis 5, 01-815 Warsaw, Poland;*
*Authors contributed equally: frank.wackenhut@uni-tuebingen.de; lukasz.j.piatkowski@put.poznan.pl

**EXPERIMENTAL SECTION**

Methyl salicylate (99%), methylcyclohexane (99%), and benzonitrile (99%) were purchased from Sigma-Aldrich and used without further purification. $CaF_2$ windows (d = 25.0 mm, 5.0 mm thickness) were purchased from Crystran LTD, while mirror mounts with piezoelectric control were purchased from Thorlabs, Inc.

**Open microcavity preparation**:[1-3] The Au films, typically 10 nm thick, were deposited on the $CaF_2$ substrate by a high vacuum sputtering (Leica EM MED 020). The thickness of the metal film was controlled using a quartz crystal balance as a reference during sputtering. The microcavity was realized by bringing the Au-coated $CaF_2$ windows to proximity and by carefully controlling their parallelism with the kinematic mirror holders (Figure 1b). Subsequently, drops of MS in methylcyclohexane with an appropriate concentration were deposited onto the edges of the $CaF_2$ windows using a micropipette until the entire microcavity was filled through capillary action.

**Optical measurements**: All transmission spectra were recorded with a Fourier transform infrared (FTIR) spectrometer (Bruker Vertex 70). Spectra were measured with a resolution of 0.5 $cm^{-1}$ and averaged over 50 subsequent scans. MS solutions (0.05% v/v, 0.5% v/v, 5% v/v, 15% v/v and 25% v/v) were prepared by dissolving MS in methylcyclohexane. The free space transmission spectra of neat MS and neat BN were obtained by injecting the liquid between two uncoated $CaF_2$ windows. Prior to vibrational strong coupling experiments, the quality of an empty cavity was verified by measuring its transmission spectrum. Then the cavity was filled with a solution and a voltage applied to the piezo element was adjusted in order to tune the cavity mode towards the molecular mode under study.

**Theoretical description:** In order to theoretically describe VSC we used up to four coupled damped harmonic oscillators, one oscillator is used to model the cavity mode and the other three oscillators are used to model the three different molecular vibrations coupling to the same cavity mode. The four equations of motion of such a coupled oscillator system can be written as follows:[4, 5]

$$\ddot{x}_1(t) + \gamma_1 \dot{x}_1(t) + \omega_1^2 x_1(t) + \kappa_2 x_2(t) + \kappa_3 x_3(t) + \kappa_4 x_4(t) = 0 \quad (1)$$

$$\ddot{x}_2(t) + \gamma_2 \dot{x}_2(t) + \omega_2^2 x_2(t) + \kappa_2 x_1(t) = 0 \quad (2)$$

$$\ddot{x}_3(t) + \gamma_3 \dot{x}_3(t) + \omega_3^2 x_3(t) + \kappa_3 x_1(t) = 0 \quad (3)$$

$$\ddot{x}_4(t) + \gamma_4 \dot{x}_4(t) + \omega_4^2 x_4(t) + \kappa_4 x_1(t) = 0 \quad (4)$$

with the damping constants $\gamma_1, \gamma_2, \gamma_3, \gamma_4$, resonance frequencies $\omega_1, \omega_2, \omega_3, \omega_4$, and coupling constants $\kappa_2, \kappa_3, \kappa_4$. Equation (1) is used to describe the cavity mode and is coupled to equations (2)-(4) via the terms proportional to $\kappa$, which allows a coherent energy exchange between the cavity mode $x_1(t)$ and the vibrations $x_{2-4}(t)$. We solved these coupled differential equations numerically to obtain the time evolution of the amplitudes $x_{1-4}(t)$ and their Fourier transforms yielded the corresponding spectrum. Especially interesting is the time evolution of the cavity mode $x_1(t)$, since it corresponds to the transmission spectrum of the coupled system and can be measured experimentally. This calculated transmission spectrum can be fitted to the experimental data and different scenarios can be evaluated, which is shown in Figure 4 of the main text. The parameters used to simulate the spectra shown in Figures 3 and 4 are given in Table 2 for benzonitrile and in Tables 3 and 4 for MS.

Table 1: Vibrational frequency peak assignments for the FTIR spectrum of MS.[6,7]

| Vibrational frequency (cm$^{-1}$) | Vibrational assignments |
|---|---|
| 3193 | O-H symmetric stretching |
| 1685 | C=O symmetric stretching |
| 1616 | Phenyl stretching vibrations |
| 1585 | |
| 1487 | |
| 1448 | |
| 1304 | |
| 1340 | In-plane O-H deformation |
| 1253 | (C=O)-O stretching |

Table 2: Parameters used to simulate spectra of benzonitrile shown in Fig.3.

| | Frequency (cm$^{-1}$) | $\gamma$ (meV) | $\kappa$ (meV) |
|---|---|---|---|
| Vibration 1 | 2229 | 1.2 | 2.3 |
| Cavity 1 | 2227 | 2.0 | - |

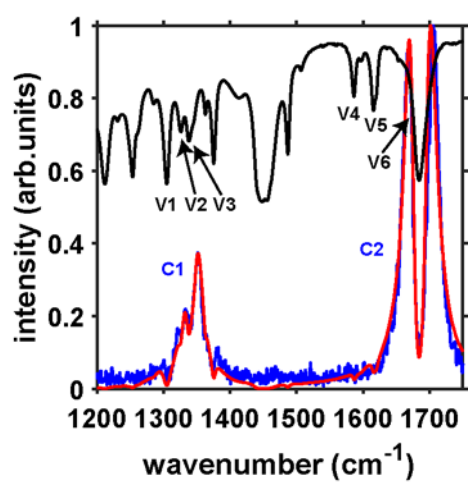

Figure S1: The blue and red lines display the experimental and calculated transmission spectra for a cavity containing 5% MS solution. The figure also shows assignment of the cavity (C1-C2) and vibrational (V1-V6) modes for the calculation of transmission spectra for the 5% MS solution. The intensity of the free space transmission spectrum of MS (black line) is scaled (x0.5) for illustration purposes.

Table 3: Parameters used to simulate spectra of 5% methyl salicylate shown in Fig. 4 (top row) and Fig. S1.

|  | Frequency (cm$^{-1}$) | $\gamma$ (meV) | $\kappa$ (meV) |
|---|---|---|---|
| Vibration 1 | 1304 | 2.0 | 0.7 |
| Vibration 2 | 1325 | 1.0 | 0.1 |
| Vibration 3 | 1340 | 2.7 | 0.6 |
| Cavity 1 | 1342 | 2.5 | - |
|  |  |  |  |
| Vibration 4 | 1586 | 0.7 | 0.5 |
| Vibration 5 | 1616 | 1.0 | 0.5 |
| Vibration 6 | 1685 | 2.4 | 1.0 |
| Cavity 2 | 1685 | 2.6 | - |

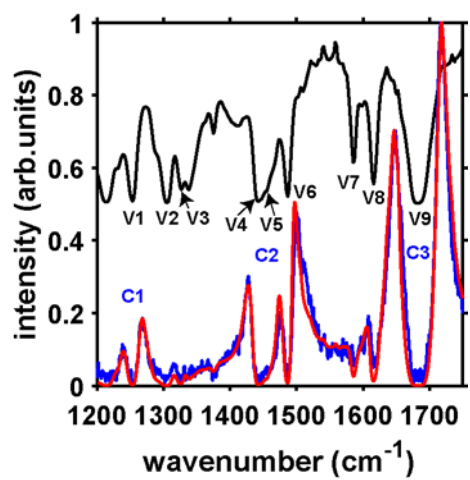

Figure S2: The blue and red lines display the experimental and calculated transmission spectra for a cavity containing 25% MS solution. The figure also shows assignment of the cavity (C1-C3) and vibrational (V1-V9) modes for the calculation of transmission spectra for the 25% MS solution. The intensity of the free space transmission spectrum of MS (black line) is scaled (x0.5) for illustration purposes.

Table 4: Parameters used to simulate spectra of 25% methyl salicylate shown in Fig. 4 (lower row) and Fig. S2.

|  | Frequency (cm$^{-1}$) | $\gamma$ (meV) | $\kappa$ |
|---|---|---|---|
| Vibration 1 | 1253 | 1.8 | 0.8 |
| Vibration 2 | 1304 | 2.0 | 1.5 |
| Vibration 3 | 1325 | 1.0 | 1.0 |
| Cavity 1 | 1275 | 2.5 | - |
|  |  |  |  |
| Vibration 4 | 1443 | 2.1 | 1.0 |
| Vibration 5 | 1458 | 3.2 | 1.1 |
| Vibration 6 | 1487 | 0.8 | 0.8 |
| Cavity 2 | 1465 | 3.3 | - |
|  |  |  |  |
| Vibration 7 | 1586 | 0.7 | 1.0 |
| Vibration 8 | 1616 | 1.0 | 1.3 |
| Vibration 9 | 1685 | 2.4 | 2.6 |
| Cavity 3 | 1674 | 4.3 | - |